\documentclass[useAMS]{mnras}

\usepackage{graphicx}
\usepackage[fleqn]{amsmath}
\usepackage{multirow}
\usepackage{hyperref}
\usepackage{xspace}
\usepackage{url}

    \renewcommand{\sun}{\hbox{$\odot$}}

    \renewcommand{\degr}{\hbox{$^\circ$}\xspace}
    \newcommand{\kms}{\hbox{km s$^{-1}$}\xspace}

    \newcommand{\lumcgs}{\hbox{erg s$^{-1}$}\xspace}
    
    \newcommand{\fek}{Fe\,K\xspace}
    \newcommand{\feka}{Fe\,K$\alpha$\xspace}
    
    \newcommand{\ka}{K$\alpha$\xspace}
    \newcommand{\kb}{K$\beta$\xspace}

    \newcommand{\fei}{Fe\,\textsc{i}\xspace}
    \newcommand{\fexxv}{Fe\,\textsc{xxv}\xspace}

    \newcommand{\dchi}{\Delta\chi^{2}}
    \newcommand{\dchidnu}{\Delta\chi^{2}/\Delta\nu}
    \newcommand{\chidof}{\chi^{2}/\nu}
    \newcommand{\redchi}{\chi^{2}_\nu}
    \newcommand{\nh}{N_{\rm H}}

    \renewcommand{\rmn}[1]{{\mathrm{#1}}}
    \newcommand{\xmm}{\emph{XMM--Newton}\xspace}
    \newcommand{\swift}{\emph{Swift}\xspace}
    
    \newcommand{\chandra}{\emph{Chandra}\xspace}
    \newcommand{\nustar}{\emph{NuSTAR}\xspace}

    \newcommand{\plcabs}{\texttt{plcabs}\xspace}
    \newcommand{\sphere}{\texttt{sphere}\xspace}
    
    \newcommand{\mytorus}{\texttt{MYTorus}\xspace}

    \newcommand{\ciao}{\textsc{ciao}\xspace}
    \newcommand{\sherpa}{\textsc{sherpa}\xspace}

    \newcommand{\matplotlib}{\texttt{matplotlib}\xspace}
    \newcommand{\python}{\textsc{python}\xspace}
    
    \newcommand{\ngc}{NGC\,6240\xspace}

\title[Hard X-ray view of \ngc]{Nuclear absorption and emission in the AGN merger \ngc: the hard X-ray view}
\author[E. Nardini]
{Emanuele Nardini$^1$\thanks{E-mail: enardini@arcetri.inaf.it} \\
$^1$INAF -- Osservatorio Astrofisico di Arcetri, Largo Enrico Fermi 5, I-50125 Firenze, Italy}

\begin{document}

\date{Released Xxxx Xxxxx XX}

\pagerange{\pageref{firstpage}--\pageref{lastpage}} \pubyear{2017}

\maketitle

\label{firstpage}

\begin{abstract}
We present the analysis of four \nustar observations of the luminous infrared 
galaxy merger \ngc, hosting a close pair of highly obscured active galactic nuclei 
(AGN). Over a period of about two years, the source exhibits hard X-ray variability 
of the order of 20 per cent, peaking around 20 keV. When the two AGN are resolved 
with \chandra, column densities in the range $\nh \sim 1$--$2 \times 10^{24}$ 
cm$^{-2}$ are estimated for both of them. The exact values are hard to determine, 
as they appear to depend on aspects that are sometimes overlooked in Compton-thick 
objects, such as the covering factor of the absorber, iron abundance, and the 
contamination in the \fek band from foreground hot-gas emission. Nearly spherical 
covering and slightly subsolar iron abundance are preferred in this case. While the 
southern nucleus is suggested to be intrinsically more powerful, as also implied by 
the mid-IR and 2--10 keV brightness ratios, solutions involving a similar X-ray 
luminosity of the two AGN cannot be ruled out. The observed variability is rather 
limited compared to the one revealed by the \swift/BAT light curve, and it can be 
fully explained by changes in the continuum flux from the two AGN, without requiring 
significant column density variations. \ngc is hereby confirmed to represent a unique 
opportunity to investigate the X-ray (and broad-band) properties of massive galaxy 
mergers, which were much more frequent in the early Universe. 
\end{abstract}


\section{Introduction}

In the framework of hierarchical structure assembly over cosmic time (e.g. Springel 
et al. 2005), major mergers are a key phase of galaxy evolution. The existence of 
close supermassive black hole (SMBH) pairs is therefore inevitable. As the merger 
advances, the loss of angular momentum from the gas and the subsequent inflows towards 
the nuclear regions trigger a violent burst of star formation and efficient accretion 
onto the SMBHs, which are then transformed into active galactic nuclei (AGN). In the 
local Universe, one of the most impressive objects of this kind is \ngc ($z \simeq 
0.0245$; Downes, Solomon \& Radford 1993). On its way to entering the ultimate stage 
of coalescence between two gas-rich spirals (Fosbury \& Wall 1979; Fried \& Schulz 
1983; Wright, Joseph \& Meikle 1984), becoming an ultra-luminous infrared galaxy 
($L_\rmn{IR} > 10^{12} L_{\sun}$) and eventually a passive elliptical (Tacconi et 
al. 1999; Bush et al. 2008), \ngc is an ideal target to look into many aspects of 
galaxy evolution at once, including merger dynamics, AGN/starburst interplay, winds 
and feedback, 
chemical enrichment. In fact, \ngc has it all: a distorted optical morphology with 
long tidal tails (Gerssen et al. 2004) and two nuclei separated by 0.7--0.8 kpc 
(Max, Canalizo \& de Vries 2007), which are likely the remnants of the original 
galactic bulges (Engel et al. 2010a); a huge ($\sim$\,10$^{10} M_{\sun}$) molecular 
gas content in the central kpc, peaking between the nuclei and pervaded by turbulence 
and shocks with different velocity (Iono et al. 2007; Meijerink et 
al. 2013; Tunnard et al. 2015); a star formation rate of several tens of solar 
masses per year (Yun \& Carilli 2002), thought to drive a starburst superwind 
that expands into the H$\alpha$ and soft X-ray emitting circumnuclear nebula 
(Heckman, Armus \& Miley 1987; Lira et al. 2002), possibly extending out to 
galactic scales (Nardini et al. 2013; Yoshida et al. 2016); and, finally, a pair 
of buried AGN, one in each nucleus, which can be clearly identified only in the 
hard X-ray band (2.5--8 keV; Komossa et al. 2003). 

As both AGN in \ngc are extremely obscured, only circumstantial evidence of their 
presence is found at other wavelengths (Gallimore \& Beswick 2004; Risaliti et al. 
2006; Armus et al. 2006; Hagiwara, Baan \& Kl{\"o}ckner 2011). Hence our knowledge 
of their energy input and of their impact on the system is still disappointingly 
scant. This largely relies upon the X-ray study, since Iwasawa \& Comastri (1998) 
detected a prominent \fek complex on top of a flat 3--10 keV continuum, resembling 
the spectrum of a typical obscured Seyfert galaxy like NGC\,1068, but nearly ten 
times more luminous. The footprint of the direct AGN emission was revealed 
soon afterwards by \textit{BeppoSAX} above 10 keV (Vignati et al. 1999). In this 
work we present the results of four \nustar observations of \ngc taken over a period 
of about two years, and discuss their implications on the intrinsic properties of 
the two AGN and on the nature of their obscuration. 

\section{Observations and Data Reduction}

\ngc was first observed by \nustar on 2014 March 30, as part of the satellite's 
Extragalactic Survey. The results of this observation have been discussed in detail 
by Puccetti et al. (2016). The source was subsequently targeted on three further 
occasions during the present monitoring programme, on 2015 April 17, 2015 September 6, 
and 2016 February 20. All the four data sets were reprocessed with the \nustar Data Analysis 
Software\footnote{\url{https://heasarc.gsfc.nasa.gov/docs/nustar/analysis/nustar_swguide.pdf}}
v1.7.1 included in the \textsc{heasoft} v6.20 release, using the 20170120 version of 
the calibration files. Cleaned events were obtained with the \texttt{nupipeline} task, 
applying the usual filtering criteria except for the more conservative (\textit{optimized}) 
treatment of the background in correspondence with the spacecraft passages through 
the South Atlantic Anomaly. This reduced the good-time interval by about 1--3 ks per 
Focal Plane Module (A and B) over each of the 2015--2016 observations. Accordingly, 
the cumulative net exposures are 92.2 ks for FPMA and 91.9 for FPMB (Table~\ref{to}). 
Source light curves and spectra were extracted from circular regions with radius of 75 
arcsec centred on the target, while the associated background data products were obtained 
from adjacent regions with radius of 90 arcsec on the same detector.\footnote{The 
differences between source and background extraction areas were properly taken into 
account at each step and for each satellite through the BACKSCAL keyword in the file 
headers.} The source counts typically exceed 90 per cent of the total counts in the 
3--78 keV band, dominating up to 50--60 keV. The spectra were rebinned in order to 
oversample the intrinsic energy resolution (400 eV below 40 keV; Harrison et al. 2013) 
by a factor of 2.5, and further grouped to ensure a significance of at least 5$\sigma$ 
per spectral channel.

Seven \xmm observations of \ngc were carried out between 2000 September 22 and 2003 
August 29, during revolutions 144, 413, 597, 599, 673, 677, and 681 (Boller et al. 2003; 
Netzer et al. 2005). All of these visits were plagued by strong background flares, so 
that only a small fraction of the potential exposure is left available after the 
application of the standard background cuts (28.0 ks for pn, 41.9 ks for MOS1, and 
43.2 ks for MOS2; Table~\ref{to}). The data were reprocessed with the Science Analysis 
System (\textsc{sas}) v16.0 and the latest calibration files. Source and background 
spectra were extracted from circular regions with radii of 30 and 60 arcsec, respectively. 
Response files were generated with the \textsc{sas} tasks \texttt{rmfgen} and 
\texttt{arfgen}, then the spectra from each detector were combined over all the epochs 
with \texttt{epicspeccombine}, and grouped to a 5$\sigma$ significance per bin. After 
verifying their consistency, the two MOS spectra were further combined into a single one. 

\begin{figure}
\includegraphics[width=8.5cm]{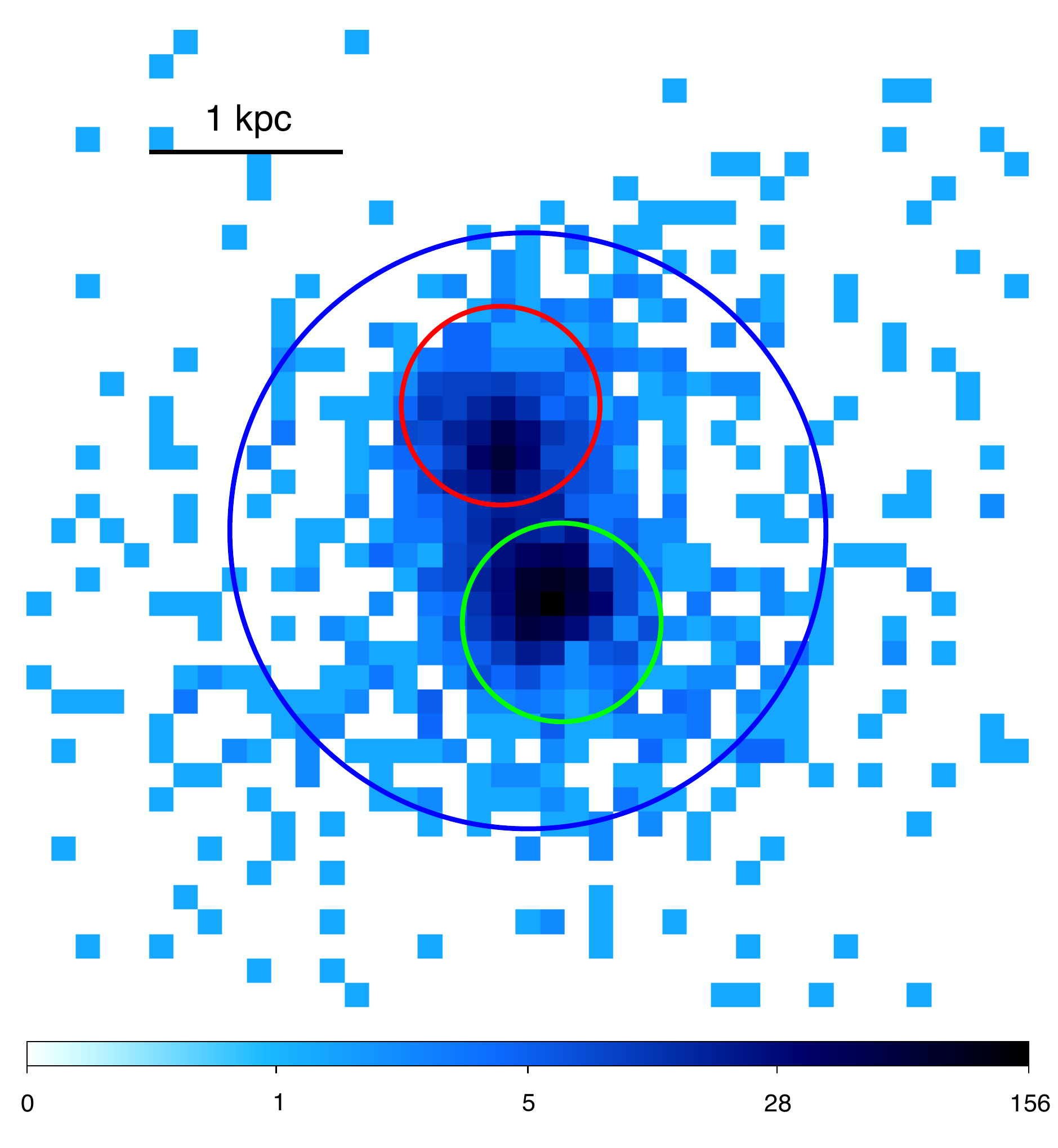}
\caption{\chandra image of the central (10$\times$10 arcsec$^2$) region of \ngc in the 
5--7.5 keV band (rest frame), obtained by combining the ACIS-S observations of 2001 July 
(ObsID 1590) and 2011 May (ObsID 12713). Subpixel resolution by a factor two is adopted, 
so one pixel in this figure has a physical size of 126~pc. The regions employed in the 
spatially resolved spectral analysis are also shown: northern nucleus (red), southern 
nucleus (green), and diffuse \fek emission (blue).} 
\label{im}
\end{figure}

\begin{table*}
\centering
\small
\caption{Observation log of the X-ray spectra analysed in this work. Net exposures and 
count rates are listed. $\mathcal{F}_\rmn{S/T}$ is the fraction of source to total counts 
in the reference band. Fluxes are derived from the best-fitting models in Section~3, and 
are not corrected for cross calibration.}
\label{to}
\begin{tabular}{l@{\hspace{20pt}}c@{\hspace{15pt}}c@{\hspace{15pt}}c@{\hspace{15pt}}c@{\hspace{15pt}}c@{\hspace{15pt}}c@{\hspace{15pt}}c}
\hline \hline
Satellite  & Date        & Detector & Exposure & Band   & Count Rate      & 
$\mathcal{F}_\rmn{S/T}$  & Flux \\
	   &             &          & (s)      & (keV)  & (s$^{-1}$)      & 
(\%) & (erg s$^{-1}$ cm$^{-2}$) \\
\hline
\nustar(1) & 2014 Mar 30 & FPMA     & 30,860   & 3--78  & 0.180$\pm$0.003 & 
92.0 & $5.08\times10^{-11}$ \\
           &             & FPMB     & 30,783   &        & 0.166$\pm$0.003 & 
89.2 & $5.02\times10^{-11}$ \\
\nustar(2) & 2015 Apr 17 & FPMA     & 20,494   & 3--78  & 0.207$\pm$0.003 & 
93.5 & $5.63\times10^{-11}$ \\
           &             & FPMB     & 20,508   &        & 0.193$\pm$0.003 & 
91.0 & $5.75\times10^{-11}$ \\
\nustar(3) & 2015 Sep 06 & FPMA     & 20,411   & 3--78  & 0.202$\pm$0.003 & 
93.7 & $5.47\times10^{-11}$ \\
           &             & FPMB     & 20,203   &        & 0.189$\pm$0.003 & 
91.0 & $5.47\times10^{-11}$ \\
\nustar(4) & 2016 Feb 20 & FPMA     & 20,421   & 3--78  & 0.180$\pm$0.003 & 
92.0 & $4.67\times10^{-11}$ \\
           &             & FPMB     & 20,433   &        & 0.168$\pm$0.003 & 
90.2 & $4.74\times10^{-11}$ \\
\xmm       & [2000--2003]  & pn       & 27,983   & 0.5--8 & 0.484$\pm$0.004 & 
97.5 & $1.98\times10^{-12}$ \\
           &             & MOS      & 85,040   &        & 0.153$\pm$0.001 & 
98.1 & $2.08\times10^{-12}$ \\
\chandra   & [2001--2011]  & ACIS-S   & 182,053  & 0.5--8 & 0.227$\pm$0.001 & 
97.1 & $2.19\times10^{-12}$ \\
\hline
\end{tabular}
\end{table*}

\ngc was also observed by \chandra with the Advanced CCD Imaging Spectrometer (ACIS) 
on four times. The observations of 2006 May 11 and 16, however, were taken with the 
High Energy Transmission Grating (HETG), thus significantly reducing the effective 
area of ACIS-S. The zeroth-order images were considered here, but they were eventually 
discarded as their quality is not sufficient to improve the results of the broad-band 
spectral analysis.\footnote{The total number of net counts in the 0.5--8 keV band 
collected in the two HETG observations is about one fourth (10500 versus 41300) of 
those collected without the gratings.} The data from the other two \chandra observations 
(performed on 2001 July 29 and 2011 May 31) were reprocessed by running the 
\texttt{chandra\underline{ }repro} script within the \textsc{ciao} v4.9 software package, 
using the Calibration Database (CALDB) version 4.7.3 and following the steps described 
in Nardini et al. (2013). With respect to the whole galaxy, the size of the regions 
employed for the spectral extraction is the same as for \xmm. Thanks to the exceptional 
angular resolution afforded by \chandra, spectra were also extracted separately for each 
nucleus and for the central (3-arcsec radius) region that encompasses, besides the nuclei, 
most of the diffuse emission in the \fek band (Wang et al. 2014). In the former case, two 
circles with radius of 1 arcsec were chosen (Fig.~\ref{im}), with a slight 
offset from the corresponding peaks of the 5--7.5 keV surface brightness in 
order to minimize the contamination from the nearby companion. Given the relative 
positions, count rates, and encircled energy fractions at 6.4 keV, this is estimated to 
be less than 2 and 4 per cent for the southern nucleus and the northern one, respectively. 
All the spectra were combined and grouped with the same criteria adopted above. 

The key details of all the data sets analysed in this work are provided in Table~\ref{to}. 
The fitting packages \textsc{xspec}\footnote{\url{https://heasarc.gsfc.nasa.gov/xanadu/xspec/}} 
v12.9.1 and \sherpa\footnote{\url{http://cxc.harvard.edu/sherpa4.9/}} v4.9 were 
used in the analysis. Unless otherwise stated, uncertainties are given at the 1$\sigma$ 
confidence level for count rates and fluxes, and at the 90 per cent level ($\dchi = 2.71$) 
for the single parameter of interest in the spectral models. The latest values of the 
concordance cosmological parameters ($H_0=67.7$ km s$^{-1}$ Mpc$^{-1}$, $\Omega_m=0.31$, 
$\Omega_\Lambda=0.69$; Planck Collaboration XIII 2016) are assumed throughout, based on 
which the luminosity distance and angular scale of \ngc are 111 Mpc and 512~pc arcsec$^{-1}$.

\section{Analysis and Discussion}

\subsection{Hard X-ray light curves}

\begin{figure}
\includegraphics[width=8.5cm]{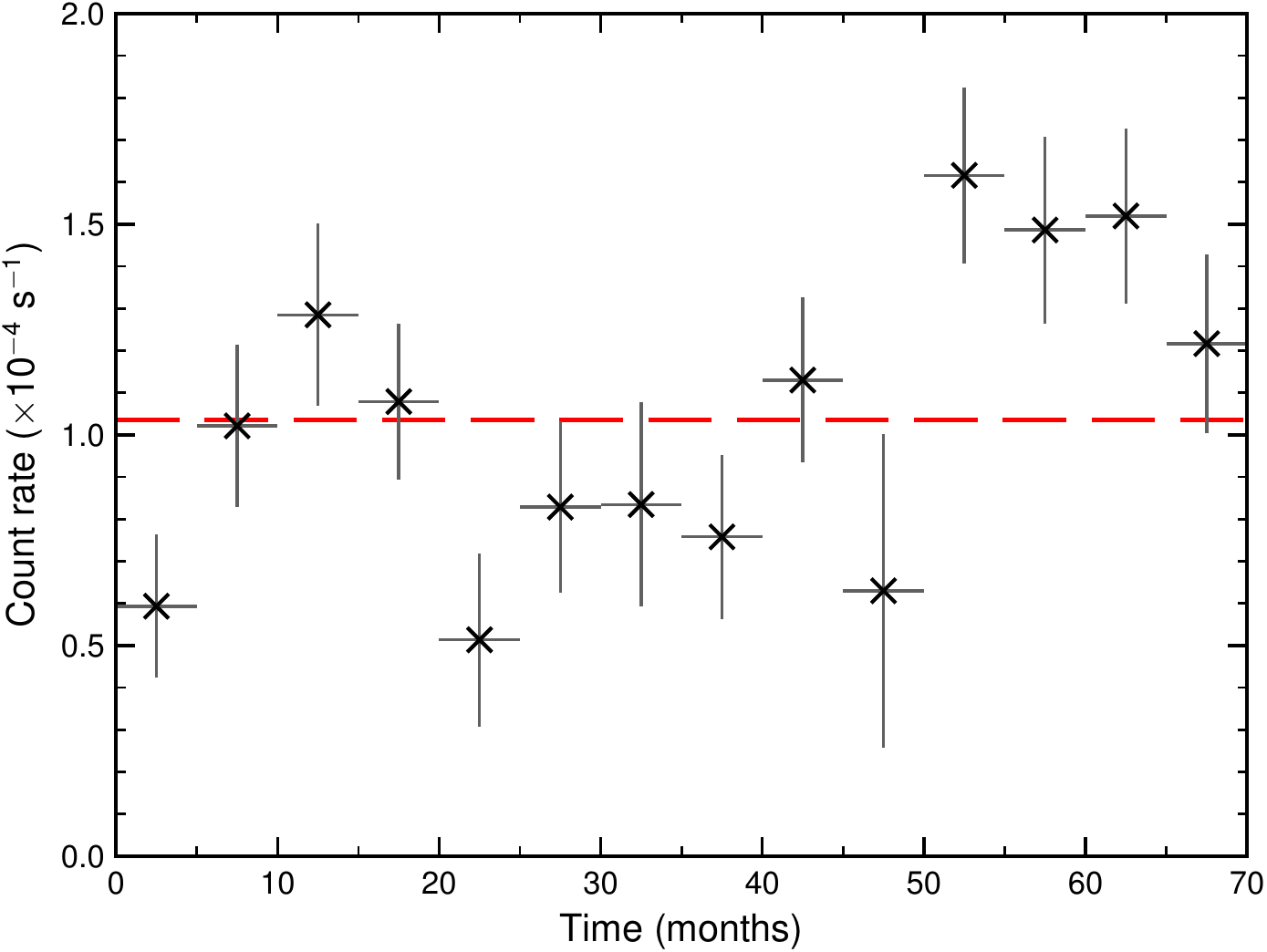}
\caption{Hard X-ray (14--195 keV) light curve of \ngc from the \swift/BAT 70-month 
survey (2004 December to 2010 September) in time bins of five months, commensurate 
with the separation between the last three \nustar observations. The source is clearly 
variable around the average (red dashed line), and significant flux changes 
can occur on time-scales of about one year.} 
\label{bs}
\end{figure}

Above 10 keV, \ngc is characterized by remarkable variability, as revealed by the 
\swift/BAT hard X-ray survey (Baumgartner et al. 2013). The 70-month 14--195 keV 
light curve, binned to periods of five months to match the time span between the 
2015--2016 \nustar observations, is shown in Fig.~\ref{bs}. The flux of the source 
is highly erratic, as it can either remain stable or vary by up to a factor of two 
over about a year. A fit with a constant returns a statistic of $\chidof = 37.8/13$, 
and can be rejected with a significance of 3.6$\sigma$. The long-term hard X-ray 
variability and its dependence on energy of more than one hundred AGN selected from 
the 58-month \swift/BAT catalogue have been recently studied by Soldi et al. (2014). 
In that work, \ngc stands out as one of the most intriguing objects. While the amplitude 
of the variability in the 14--24 and 35--100 keV bands is strongly correlated over the 
entire sample, in \ngc the overall intensity changes (e.g. Fig.~\ref{bs}) appear to be 
almost exclusively due to the lower energy band. 

Since the column density towards the nuclear regions of \ngc lies most likely in the 
range 1--$2 \times 10^{24}$ cm$^{-2}$ (Vignati et al. 1999; Ikebe et al. 2000; Puccetti 
et al. 2016), that is, around or just above the conventional threshold for Compton 
thickness,\footnote{This is defined as $\nh > (x\sigma_\rmn{T})^{-1} \sim 1.23 \times 
10^{24}$ cm$^{-2}$, where $x$ is the mean number of electrons per hydrogen atom and 
$\sigma_\rmn{T}$ is the Thomson cross section (e.g. Murphy \& Yaqoob 2009).} any change 
in $\nh$ of the order of a few $\times$10$^{23}$cm$^{-2}$ could be readily detectable 
at 5--10 keV. The historical spectra, however, are virtually consistent with each other 
over this range, within the cross calibration uncertainties between the various X-ray 
observatories (although it should be noted that the quality of the individual spectra, 
other than the 2011 \chandra ACIS-S one, is generally too low to derive robust constraints 
in this sense). The peak of the variability around 20 keV therefore remains somewhat 
puzzling. With its imaging capabilities that allow for the first time an accurate 
subtraction of the hard X-ray background, \nustar can finally shed new light on the 
striking behaviour of \ngc at high energies.

\begin{figure}
\includegraphics[width=8.5cm]{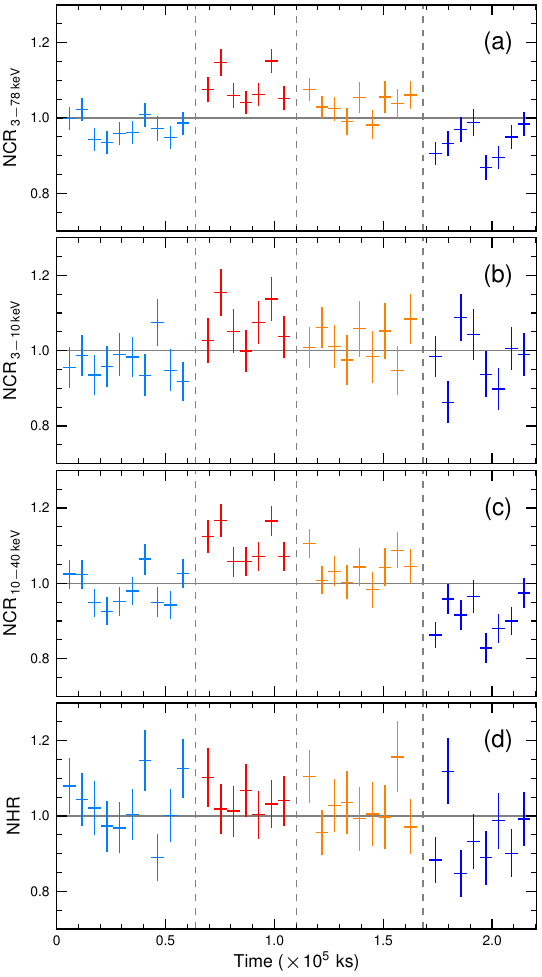}
\caption{From top to bottom: normalized count rates in the (a) 3--78, (b) 3--10, (c) 
10--40 keV bands, and (d) normalized 10--$40/3$--10 keV hardness ratio for the four 
\nustar observations. The background-subtracted light curves from both modules were 
added together, after the standard corrections for dead time, PSF, and vignetting 
losses. The horizontal scale portrays the total elapsed time, with the gaps between 
the observations removed.}
\label{lc}
\end{figure}

The 3--78 keV light curves from the four \nustar observations, normalized 
to the average count rate, are plotted in Fig.~\ref{lc} (top panel). Absolute values 
are provided in Table~\ref{tt}, together with the results obtained by fitting each 
sequence and the overall trend with a constant. A time bin of 5800 s has been adopted, 
roughly corresponding to the satellite's orbital period. The final bin of the first 
sequence was discarded because of its low ($\sim$\,10 per cent) fractional exposure. 
Over the entire \nustar band, the source is clearly variable between the different 
epochs. In the brightest state (2015 April) the count rate is about 15 per cent higher 
than in the faintest one (2016 February). Even though this range is not as large as 
that exhibited in the whole \swift/BAT survey, the statistical significance of the 
variability increases to the 7.5$\sigma$ level. Conversely, the short-term fluctuations 
noticed within the single observations can be in the first instance neglected, as the 
null-hypothesis probability of the constant fits is always $>$\,0.05 (Table~\ref{tt}). 

Light curves were also extracted separately in the soft (3--10 keV) and hard (10--40 keV) 
\nustar bands, and the hardness ratio of the latter over the former has been computed 
(Fig.~\ref{lc}). This brings out the energy dependence of the above variability, which 
is again confirmed to be more pronounced around 20 keV. Indeed, the 3--10 keV count rate 
is broadly consistent with being constant ($\chidof = 41.7/33$), and the global intensity 
changes between the observations are mostly driven by the 10--40 keV band ($\chidof = 
138/33$). In turn, this implies that some spectral variations are in place, most likely 
associated with the direct continuum level (see below). The 10--$40/3$--10 keV hardness 
ratio bears some interesting indications as well. In spite of 3--78 keV count rates that 
differ by up to $\sim$\,10 per cent, the first three sequences have almost identical 
hardness ratios, while the last sequence is not only the faintest, but also the softest 
(Table~\ref{tt}). The occurrence of states with similar (different) flux but different 
(similar) spectral shape suggests that the high-energy variability of \ngc might have 
a complex origin, possibly involving a contribution from both nuclei. 

\begin{table*}
\centering
\small
\caption{Average values of the 3--78, 3--10, 10--40 keV net count rates (from both FPMA 
and FPMB) and of the 10--$40/3$--10 keV hardness ratio for each \nustar observation, with 
fit statistic and null-hypothesis probability relative to a constant model.}
\label{tt}
\begin{tabular}{l@{\hspace{25pt}}c@{\hspace{20pt}}c@{\hspace{20pt}}c@{\hspace{20pt}}c@{\hspace{20pt}}c}
\hline \hline
Observation & 2014 Mar & 2015 Apr & 2015 Sep & 2016 Feb & All \\
\hline
$\langle$CR$_\rmn{3-78\,keV}$$\rangle$ (s$^{-1}$) & 0.543$\pm$0.005 & 0.604$\pm$0.007 & 
0.578$\pm$0.007 & 0.521$\pm$0.006 & 0.558$\pm$0.003 \\
$\chidof$ & 8.2/9 & 11.5/6 & 6.2/8 & 12.3/7 & 132/33 \\
Probability & 0.512 & 0.074 & 0.629 & 0.091 & 8.7$\times$10$^{-14}$ \\[1ex]   
$\langle$CR$_\rmn{3-10\,keV}$$\rangle$ (s$^{-1}$) & 0.175$\pm$0.003 & 0.193$\pm$0.004 & 
0.185$\pm$0.004 & 0.176$\pm$0.004 & 0.181$\pm$0.002 \\
$\chidof$ & 5.2/9 & 5.8/6 & 4.0/8 & 10.9/7 & 41.7/33 \\
Probability & 0.817 & 0.450 & 0.853 & 0.143 & 0.142 \\[1ex]
$\langle$CR$_\rmn{10-40\,keV}$$\rangle$ (s$^{-1}$) & 0.368$\pm$0.005 & 0.412$\pm$0.006 & 
0.389$\pm$0.005 & 0.340$\pm$0.005 & 0.375$\pm$0.003 \\
$\chidof$ & 13.3/9 & 8.7/6 & 6.7/8 & 12.3/7 & 138/33 \\
Probability & 0.150 & 0.190 & 0.566 & 0.091 & 8.0$\times$10$^{-15}$ \\[1ex]
$\langle$HR$\rmn{\frac{10-40\,keV}{3-10\,keV}}$$\rangle$ & 2.087$\pm$0.046 & 2.129$\pm$0.053 & 
2.096$\pm$0.053 & 1.908$\pm$0.050 & 2.054$\pm$0.025 \\
$\chidof$ & 10.5/9 & 1.4/6 & 5.3/8 & 9.0/7 & 37.8/33 \\
Probability & 0.308 & 0.967 & 0.722 & 0.254 & 0.258 \\[1ex]
\hline
\end{tabular}
\end{table*}

\subsection{Average \nustar spectrum}

While there is tentative evidence that the fourth observation caught \ngc in a `softer' 
state, the hardness ratio is formally consistent with a constant value ($\langle$HR$\rangle 
\simeq 2.05$; $\chidof = 37.8/33$), i.e. with no evolution with time. The limited extent 
of any spectral variability then allows us to combine all the data from the four epochs, 
and to take advantage of the very high quality of the resultant \nustar spectrum (adding 
up to a total exposure of $\sim$\,184 ks between FPMA and FPMB) to derive some preliminary 
information on the properties of the obscuring medium. The soft X-ray emission is probed 
through the \xmm (pn and MOS) and \chandra (ACIS-S) spectra, which also provide a better 
energy resolution in the \fek band and a benchmark against variability below 10 keV. For 
ease of comparison, the \xmm and \chandra spectra were fitted over the common range of 
0.5--8 keV, outside of which the effective area of the MOS and ACIS-S detectors rapidly 
drops. The individual \nustar spectra from FPMA and FPMB were analysed in the 3--78 keV 
band. 

Thanks to the vast knowledge on this system accumulated with the previous X-ray studies, 
here we can directly move to a physically motivated picture of the thermal, diffuse 
emission and of the obscured, nuclear components. The contribution from collisionally 
ionized gas with at least three different temperatures has been identified in \ngc (Boller 
et al. 2003). The gas becomes gradually hotter (and more obscured) as one moves towards 
the centre, from the cold ($kT \sim 0.6$--0.7 keV) halo detected out to scales of about 
100 kpc (Nardini et al. 2013), through the warm ($kT \sim 1$--1.5 keV) butterfly-shaped 
circumnuclear nebula (Lira et al. 2002), to the shock-heated ($kT \sim 5$--6 keV) 
environment around and between the nuclei (Wang et al. 2014). With this in mind, the 
initial spectral model has been defined as follows within \textsc{xspec}:
\begin{center}
Model $\mathcal{A}$ = \texttt{constant[1]}$\times$\texttt{phabs[2]}$\times$(\texttt{vmekal[3]} + 
\texttt{zphabs[4]}$\times$\texttt{mekal[5]} + \texttt{zphabs[6]}$\times$\texttt{pshock[7]} + 
\texttt{MYTZ[8]}$\times$\texttt{zpowerlw[9]} + \texttt{constant[10]}$\times$\texttt{MYTS[11]} 
+ \texttt{constant[12]}$\times$\texttt{MYTL[13]} + 
\texttt{zphabs[14]}$\times$\texttt{zpowerlw[15]}),
\end{center}
where \texttt{vmekal[3]}, \texttt{mekal[5]}, and \texttt{pshock[7]} represent the thermal 
emission from the cold, warm, and hot/shocked plasma, respectively. Abundances of iron and 
$\alpha$-elements were allowed to vary in the first component, while solar values (from 
Anders \& Grevesse 1989) were assumed in the other two, which are also affected by foreground 
absorption (\texttt{zphabs[4,6]}), presumably associated with the cold halo itself and/or 
with the prominent galactic dust lanes (e.g. Max et al. 2005). Compton-thick obscuration and 
reprocessing of the intrinsic nuclear continuum (\texttt{zpowerlw[9]}) were implemented with 
the \mytorus model (Murphy \& Yaqoob 2009), which self-consistently computes the line-of-sight 
attenuation (\texttt{MYTZ[8]}), the scattered continuum (\texttt{MYTS[11]}), and the 
fluorescent emission lines (\texttt{MYTL[13]}) from a toroidal absorber with circular cross 
section and half-opening angle of 60\degr. The \texttt{constant[10,12]} factors adjust the 
relative weights of the three components, and were initially fixed to 1 (as per the standard 
`coupled' configuration; Yaqoob 2012). Given the geometry of \mytorus, a fraction of the 
direct continuum (\texttt{zpowerlw[15]}) can be scattered into the line of sight by optically 
thin ionized gas in the polar regions,\footnote{This is referred to as `reflected continuum' 
in the following, in order to avoid confusion with the scattered component from the cold, 
Compton-thick material.} before being possibly absorbed as well (\texttt{zphabs[14]}). 
The Galactic column density (\texttt{phabs[2]}) was set to $4.87 \times 10^{20}$ cm$^{-2}$ 
(Kalberla et al. 2005), while \texttt{constant[1]} accounts for the cross normalization 
between the spectra from different detectors, compared to FPMA. 

When applied to the \xmm, \chandra, and average \nustar spectra, Model $\mathcal{A}$ gives 
a fairly good fit statistic ($\chidof \simeq 1482/1310$). The equatorial column density of 
the torus is $1.3 \times 10^{24}$ cm$^{-2}$, and its inclination angle is 87\degr. The 
direct continuum dominates at all energies over the scattered one. Yet this model is not 
acceptable, due to the presence of clear residuals in the 6--10 keV range. These take the 
shape of a modest excess, to compensate for which the shock component tends to a much 
higher temperature than previously reported, $kT = 9.1^{+2.7}_{-1.4}$ keV. Moreover, the 
strength of the \fei and \fexxv lines is not perfectly reproduced. Such a defective outcome 
is not particularly surprising, given the rigidity of the basic \mytorus configuration as 
adopted in Model $\mathcal{A}$. These assumptions can be relaxed by `decoupling' the three 
\mytorus tables to various degrees (see Yaqoob 2012 for an extensive review). For instance, 
the relative weights \texttt{[10,12]} can be allowed to vary to introduce the effects of time 
delays between the direct and the reprocessed emission, and the column densities along the 
line of sight and outside of it can be different. More complex set-ups mimic a clumpy medium. 
However, none of these solutions works here. Nor does the inclusion in the model of a second 
nucleus, by doubling the components \texttt{[8]} to \texttt{[13]}. The improvement is always 
marginal, leaving a $\redchi > 1.12$. The actual shortcomings with Model $\mathcal{A}$ must 
then reside in one of the features of \mytorus that cannot be circumvented through an ad-hoc 
decoupling, and specifically the fixed abundances and the global covering factor. Within 
\mytorus, it is not trivial to obtain a covering factor much larger than the native 
$\Omega/4\upi = 0.5$, and also the assumption of solar abundances (especially for iron) 
can have a substantial impact when the data quality is very high. 

Thanks to the growing interest in the properties of the Compton-thick AGN population, in 
the last years several other models have been presented that accurately describe absorption, 
scattering, and fluorescence in an $\nh > 10^{24}$ cm$^{-2}$ medium (e.g. Ikeda, Awaki \& 
Terashima 2009; Brightman \& Nandra 2011; Liu \& Li 2014). Among these, the only geometry 
that can help us overcome the aforementioned limitations of \mytorus is the uniform 
distribution of matter over the entire solid angle ($\Omega/4\upi = 1$) of Brightman \& 
Nandra (2011), whose spectral counterpart is dubbed for simplicity as \sphere from now on. 
In fact, a further advantage of this model is that iron and total (i.e. common to all the 
other elements) abundances are variable. On the other hand, the various contributions from 
transmission, scattering, and fluorescence cannot be separated. Thus in Model $\mathcal{B}$ 
the former components \texttt{[8]} to \texttt{[13]} were replaced by a single \sphere 
spectral table. The absorbed reflected power law was retained, as this barely compromises 
the physical and geometrical self-consistency of this picture (see also LaMassa et al. 2014, 
and below). With abundances frozen to solar, it turns out that Model $\mathcal{B}$ ($\chidof 
\simeq 1508/1311$) is even worse than Model $\mathcal{A}$, as the Compton shoulder of the 
neutral \feka line (e.g. Yaqoob \& Murphy 2011) is slightly overestimated and adds to the 
previous residuals. Quite interestingly, the shock temperature is now pushed to the small 
value of $kT = 3.2^{+0.7}_{-0.4}$ keV, due to the increased amount of scattering into 
the line of sight. A complete covering thus appears to be not necessarily involved, 
justifying some reflection of the intrinsic continuum that leaks along the clear 
directions.

\begin{figure}
\includegraphics[width=8.5cm]{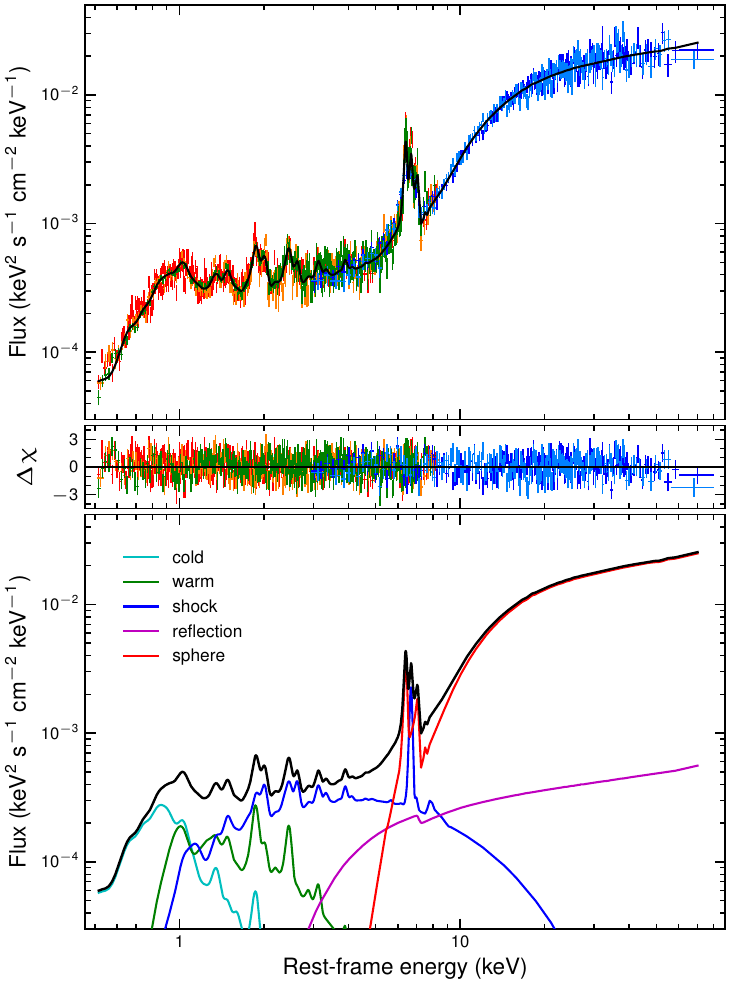}
\caption{Top panel: average \xmm (red for pn, orange for MOS), \chandra (green), and \nustar 
spectra (light and deep blue for FPMA and FPMB) of \ngc, corrected for cross normalization. 
Best-fitting Model $\mathcal{B}$ is superimposed (black curve). Mid panel: residuals in units 
of $\sigma$. Bottom panel: contribution of the different spectral components, convolved with 
the \chandra ACIS-S response up to 8 keV, and with the \nustar one beyond.}
\label{sf}
\end{figure}

\begin{table}
\centering
\small
\caption{Best-fitting parameters for Model $\mathcal{B}$ and Model $\mathcal{C}$ (see text 
for details), applied to the average \xmm, \chandra, and \nustar spectra. Galactic absorption 
is frozen at $4.87 \times 10^{20}$ cm$^{-2}$. Abundances are taken to be solar when not 
specified. Continuum normalizations ($K$) are computed at 1 keV. EM $= \int{n_e n_\rmn{H} 
\rmn{d}V}$ (where $n_e$ and $n_\rmn{H}$ are the electron and hydrogen densities) is the 
emission measure of the gas. $\tau_\rmn{ion}$ is the ionization time-scale (see Smith \& 
Hughes 2010). (f) and (t) denote frozen and tied parameters, respectively. The following 
cross normalization factors were obtained with respect to FPMA: 0.85$\pm$0.03 (pn), 
0.89$\pm$0.03 (MOS), 0.94$\pm$0.03 (ACIS-S), and 1.01$\pm$0.02 (FPMB).}
\label{tf}
\begin{tabular}{l@{\hspace{15pt}}c@{\hspace{10pt}}c}
\hline \hline
Component & \multirow{2}{*}{Model $\mathcal{B}$} & \multirow{2}{*}{Model $\mathcal{C}$} \\
~~Parameter & & \\
\hline
\multicolumn{3}{l}{\texttt{vmekal} (cold thermal emission)} \\[1ex]
~~$kT$ (keV) & 0.56$^{+0.02}_{-0.04}$ & 0.56$^{+0.02}_{-0.04}$ \\[0.8ex] 
~~$Z_\alpha$ (solar) & 0.40$^{+0.13}_{-0.09}$ & 0.40$^{+0.13}_{-0.10}$ \\[0.8ex] 
~~$Z_\rmn{Fe}$ (solar) & 0.19$^{+0.04}_{-0.03}$ & 0.19$^{+0.05}_{-0.03}$ \\[0.8ex] 
~~EM (10$^{63}$ cm$^{-3}$) & 54$^{+8}_{-9}$ & 53$^{+9}_{-8}$ \\[0.8ex] 
\multicolumn{3}{l}{\texttt{mekal} (warm thermal emission)} \\[1ex]
~~$\nh$ (10$^{22}$ cm$^{-2}$) & 0.95$\pm$0.04 & 0.95$\pm$0.04 \\[0.8ex] 
~~$kT$ (keV) & 0.84$^{+0.04}_{-0.05}$ & 0.84$\pm$0.04 \\[0.8ex] 
~~EM (10$^{63}$ cm$^{-3}$) & 119$^{+18}_{-16}$ & 119$\pm$17 \\[0.8ex] 
\multicolumn{3}{l}{\texttt{pshock} (hot thermal emission)} \\[1ex]
~~$\nh$ (10$^{22}$ cm$^{-2}$) & 0.95(t) & 0.95(t) \\[0.8ex]  
~~$kT$ (keV) & 5.1$^{+1.1}_{-0.8}$ & 5.0$^{+1.1}_{-0.8}$ \\[0.8ex] 
~~$\tau_\rmn{ion}$ (10$^{11}$ s cm$^{-3}$) & 5.0$^{+1.6}_{-1.2}$ & 5.2$^{+1.9}_{-1.2}$ \\[0.8ex] 
~~EM (10$^{63}$ cm$^{-3}$) & 108$\pm$9 & 108$\pm$9 \\[0.8ex] 
\multicolumn{3}{l}{\sphere/\plcabs (nuclear emission)} \\[1ex]	
~~$\Gamma$ & 1.72$^{+0.04}_{-0.06}$ & 1.82$\pm$0.05 \\[0.8ex] 
~~$\nh$ (10$^{22}$ cm$^{-2}$) & 129$^{+5}_{-6}$ & 156$^{+19}_{-9}$ \\[0.8ex] 
~~$Z_\rmn{Fe}$ (solar) & 0.72$\pm$0.03 & 0.79$^{+0.10}_{-0.17}$ \\[0.8ex] 
~~$K$ (10$^{-2}$ s$^{-1}$ cm$^{-2}$ keV$^{-1}$) & 0.68$^{+0.11}_{-0.12}$ & 
1.07$^{+0.22}_{-0.16}$ \\[0.8ex] 
\multicolumn{3}{l}{\texttt{2\,zgauss} (fluorescent lines)} \\[1ex]
~~$E_1$ (keV) & $-$ & 6.391$^{+0.006}_{-0.007}$ \\[0.8ex]
~~$A_1$ (10$^{-5}$ s$^{-1}$ cm$^{-2}$) & $-$ & 1.46$\pm$0.11 \\[0.8ex]
~~$E_2$ (keV) & $-$ & 7.017$^{+0.035}_{-0.017}$ \\[0.8ex]
~~$A_2$ (10$^{-5}$ s$^{-1}$ cm$^{-2}$) & $-$ & 0.22$\pm$0.08 \\[0.8ex]
\multicolumn{3}{l}{\texttt{zpowerlw} (reflected continuum)} \\[1ex]
~~$\nh$ (10$^{22}$ cm$^{-2}$) & 12$\pm$4 & 15$^{+8}_{-4}$ \\[0.8ex]
~~$\Gamma$ & 1.72(t) & 1.82(t) \\[0.8ex] 
~~$K$ (10$^{-4}$ s$^{-1}$ cm$^{-2}$ keV$^{-1}$) & 1.65$^{+0.43}_{-0.41}$ & 
2.52$^{+1.17}_{-0.59}$ \\[2ex]
$\chidof$ & 1357.5/1310 & 1352.0/1306 \\[0.8ex]
Probability & 0.176 & 0.183 \\
\hline
\end{tabular}
\end{table}

As a further step, iron abundance was allowed to vary. This eventually delivers a very good 
fit ($\chidof \simeq 1358/1310$) with no significant residuals (Fig.~\ref{sf}). A sensible 
temperature of $kT \sim 5$\,($\pm$1) keV is recovered for the shocked plasma, while the 
properties of the cold gas phase are in good agreement with those of the extended soft X-ray 
halo (Nardini et al. 2013). Temperature and obscuration ($\nh \sim 10^{22}$ cm$^{-2}$) of 
the warm component are consistent with those found by Puccetti et al. (2016), who used a 
similar model for the diffuse thermal emission. The best-fitting \sphere parameters, 
however, are somewhat unexpected. The column density and the power-law photon index (see 
Table~\ref{tf}) are the same previously obtained with Model $\mathcal{A}$, but the intensity 
of the primary continuum is nearly twice as low. A higher covering factor naturally implies 
a larger contribution from scattering, hence the same hard X-ray flux can be produced by a 
fainter source. The direct and scattered components cannot be distinguished from each other 
within \sphere, so their relative importance cannot be immediately quantified. This 
notwithstanding, the drop in the continuum intensity is possibly too large, also given the 
subsolar iron abundance ($Z_\rmn{Fe} \simeq 0.7$) required to achieve a good fit. 
 
The toroidal version of the Brightman \& Nandra (2011) models has been claimed to overestimate 
the strength of the reprocessed (scattered plus fluorescent) emission (see Liu \& Li 2015). 
To verify if this is could be the case also for the \sphere component employed in 
Model $\mathcal{B}$, we switched to \plcabs (Yaqoob 1997), which is based on the same 
assumptions (an isotropic source located at the centre of a uniform, spherical distribution 
of matter) and still allows for variable iron abundance. As \plcabs does not account 
for line fluorescence,\footnote{It does include, instead, continuum scattering into the 
line of sight, so that any additional such component is not required.} two 
\texttt{zgauss} profiles were added in the new Model $\mathcal{C}$ for the \fei \ka and 
\kb. Moreover, the non-relativistic approximations become less accurate above 20 keV (see 
Yaqoob 1997 for details). Keeping in mind that the output of both models should be treated 
with some caution, \plcabs provides nonetheless a direct, meaningful comparison 
with \sphere. The maximum number of scatterings and the critical albedo were set 
to 10 and 0.1, respectively. On statistical grounds, the use of \plcabs is fully 
equivalent to that of \sphere, with $\chidof \simeq 1352/1306$ and all the parameters 
of the thermal emission that are perfectly matched between the two models (Table~\ref{tf}). 
Also the properties of the reflected continuum are in excellent agreement. In both cases 
the scattering fraction is $\sim$\,2--3 per cent, as usually found in obscured AGN (Noguchi 
et al. 2010), and the foreground absorption is of the order of $\sim$\,10$^{23}$ cm$^{-2}$, 
a typical obscuration for the central regions of \ngc (Iono et al. 2007). 

All of the other quantities associated with the nuclear component are moderately different. 
The continuum intensity is increased (close to and consistent with the one inferred from 
\mytorus), although the photon index is now steeper ($\Gamma \simeq 1.8$ versus 1.7) and 
the column density is larger by about 20 per cent. Subsolar iron abundance ($Z_\rmn{Fe} 
\simeq 0.8$) is preferred in Model $\mathcal{C}$ too, but a solar value cannot be 
conclusively rejected ($\chidof \simeq 1360/1307$ for $Z_\rmn{Fe} \equiv 1$). Based on 
these results, it seems plausible that the amount of scattering in \sphere is indeed 
overestimated. The discrepancies, however, are not severe. Taking the intrinsic 2--10 keV 
flux as a reference, \sphere yields a continuum fainter by 25--30 per cent with respect 
to \plcabs, and a smaller difference applies to the column densities. It is not the 
scope of this work to further investigate the performances of the various models, also 
because \ngc is not the ideal target for such a study. In the following we thus stick to 
the use of Model $\mathcal{B}$, as \plcabs is rather time-consuming with the adopted 
computational settings and does not treat fluorescence in a self-consistent way.

\begin{figure}
\includegraphics[width=8.5cm]{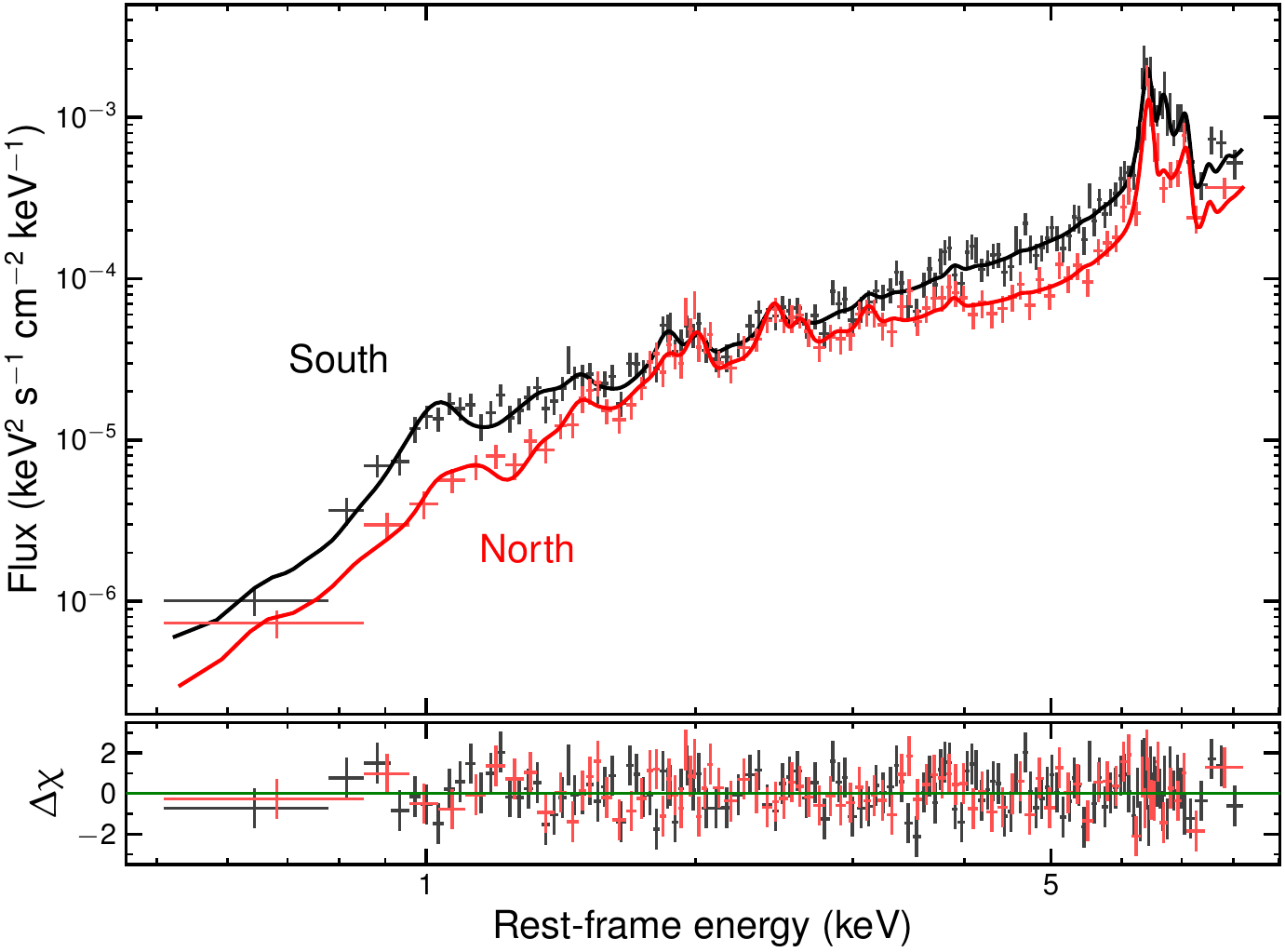}
\caption{Resolved \chandra spectra of the southern (black) and northern (red) nucleus of \ngc, 
with best-fitting models and relative residuals (in units of $\sigma$). Even the simple visual 
inspection reveals that all of the thermal components are inherently clumpy, and that their 
temperature and obscuration is locally variable. Note that some additional Ni\,\ka emission 
at 7.47 keV might be present in both nuclei.}
\label{rn}
\end{figure}

\subsection{The `double-nucleus' issue}

Differently from other bright, heavily obscured AGN with very high-quality \nustar spectra 
that show significant hard X-ray variability (e.g. NGC\,4945, Puccetti et al. 2014; NGC\,1068, 
Marinucci et al. 2016; Mrk\,3, Guainazzi et al. 2016), the interpretation of \ngc is further 
complicated by its `double-nucleus' nature. None of the previous models can be regarded as an 
adequate representation of the actual X-ray illumination and reprocessing pattern in \ngc, 
although 
the inclusion of a second source is not statistically required. A first attempt to assess the 
contribution to the hard X-ray emission from each nucleus was made by Puccetti et al. (2016), 
by fitting the spatially resolved \chandra spectra and extrapolating to higher energies. Here 
we revisit this approach in the light of the indications of the time-averaged analysis above. 
The spectra of the two AGN are shown in Fig.~\ref{rn}, and their visual inspection is already 
highly informative. The northern nucleus is known to be fainter, not only in the X-rays but 
also in the near- and mid-infrared (Risaliti et al. 2006; Asmus et al. 2014; Mori et al. 2014). 
The north/south brightness ratio is highest at 2--10 keV (about 0.6), and it can be still 
higher when referred to the AGN only. Indeed, the contamination from the hot plasma is 
much heavier in the southern nucleus, as revealed by the strength of the \fexxv emission line, 
while there is no compelling evidence for such a prominent feature in the northern nucleus. 
The two spectra are intriguing also at lower energies. Using Fig.~\ref{sf} as a guidance, 
a major contribution to the Si\,\textsc{xiii}--\textsc{xiv} and S\,\textsc{xv}--\textsc{xvi} 
complexes at 1.8--2.6 keV is expected from the shocked gas. Their nearly identical intensity 
might be accidental though, since the presumed dearth of hot matter around the northern nucleus 
would entail enhanced emission from the warm component. The spectra depart again from each 
other below 1.5 keV, hinting at differential obscuration and/or inhomogeneity in the cold gas 
phase. 

\begin{table}
\centering
\small
\caption{Best-fitting parameters for the spatially-resolved \chandra spectra of the two 
nuclei, supplemented by the average 15--78 keV \nustar spectra. The definitions are the 
same of Table~\ref{tf}.}
\label{tn}
\begin{tabular}{l@{\hspace{15pt}}c@{\hspace{10pt}}c}
\hline \hline
Component & \multirow{2}{*}{North} & \multirow{2}{*}{South} \\
~~Parameter & & \\
\hline
\multicolumn{3}{l}{\texttt{vmekal} (cold thermal emission)} \\[1ex]
~~$kT$ (keV) & 0.56(f) & 0.56(f) \\[0.8ex] 
~~$Z_\alpha$ (solar) & 0.40(f) & 0.40(f) \\[0.8ex] 
~~$Z_\rmn{Fe}$ (solar) & 0.19(f) & 0.19(f) \\[0.8ex] 
~~EM (10$^{63}$ cm$^{-3}$) & $<$\,0.29 & 0.43$\pm$0.15 \\[0.8ex] 
\multicolumn{3}{l}{\texttt{mekal} (warm thermal emission)} \\[1ex]
~~$\nh$ (10$^{22}$ cm$^{-2}$) & 3.0$^{+1.2}_{-0.7}$ & 1.2$\pm$0.2 \\[0.8ex] 
~~$kT$ (keV) & 1.47$^{+0.36}_{-0.30}$ & 1.07$^{+0.20}_{-0.09}$ \\[0.8ex] 
~~EM (10$^{63}$ cm$^{-3}$) & 37$\pm$13 & 18$\pm$7 \\[0.8ex] 
\multicolumn{3}{l}{\texttt{pshock} (hot thermal emission)} \\[1ex]
~~$\nh$ (10$^{22}$ cm$^{-2}$) & 0.8$^{+0.5}_{-0.4}$ & 19$^{+15}_{-8}$ \\[0.8ex]  
~~$kT$ (keV) & 5.1(f) & 5.1(f) \\[0.8ex] 
~~$\tau_\rmn{ion}$ (10$^{11}$ s cm$^{-3}$) & 5.0(f) & 5.0(f) \\[0.8ex] 
~~EM (10$^{63}$ cm$^{-3}$) & 3.6$^{+5.5}_{-2.6}$ & 58$^{+22}_{-14}$ \\[0.8ex] 
\multicolumn{3}{l}{\sphere (nuclear emission)} \\[1ex]	
~~$\Gamma$ & 1.78$^{+0.09}_{-0.04}$ & 1.78(t) \\[0.8ex] 
~~$\nh$ (10$^{22}$ cm$^{-2}$) & 153$^{+40}_{-18}$ & 143$^{+12}_{-27}$ \\[0.8ex] 
~~$Z_\rmn{Fe}$ (solar) & 0.70$^{+0.10}_{-0.08}$ & 0.70(t) \\[0.8ex] 
~~$K$ (10$^{-2}$ s$^{-1}$ cm$^{-2}$ keV$^{-1}$) & 0.37$^{+0.32}_{-0.11}$ & 
0.49$^{+0.13}_{-0.15}$ \\[0.8ex] 
\multicolumn{3}{l}{\texttt{zpowerlw} (reflected continuum)} \\[1ex]
~~$\nh$ (10$^{22}$ cm$^{-2}$) & 14$^{+13}_{-7}$ & 3.0$^{+2.0}_{-1.5}$ \\[0.8ex]
~~$\Gamma$ & 1.78(t) & 1.78(t) \\[0.8ex] 
~~$K$ (10$^{-4}$ s$^{-1}$ cm$^{-2}$ keV$^{-1}$) & 0.71$^{+0.52}_{-0.24}$ & 
0.59$^{+0.33}_{-0.27}$  \\[2ex]
$\chidof$ & \multicolumn{2}{c}{414.9/432} \\[0.8ex]
Probability & \multicolumn{2}{c}{0.715} \\
\hline
\end{tabular}
\end{table}

\begin{figure}
\includegraphics[width=8.5cm]{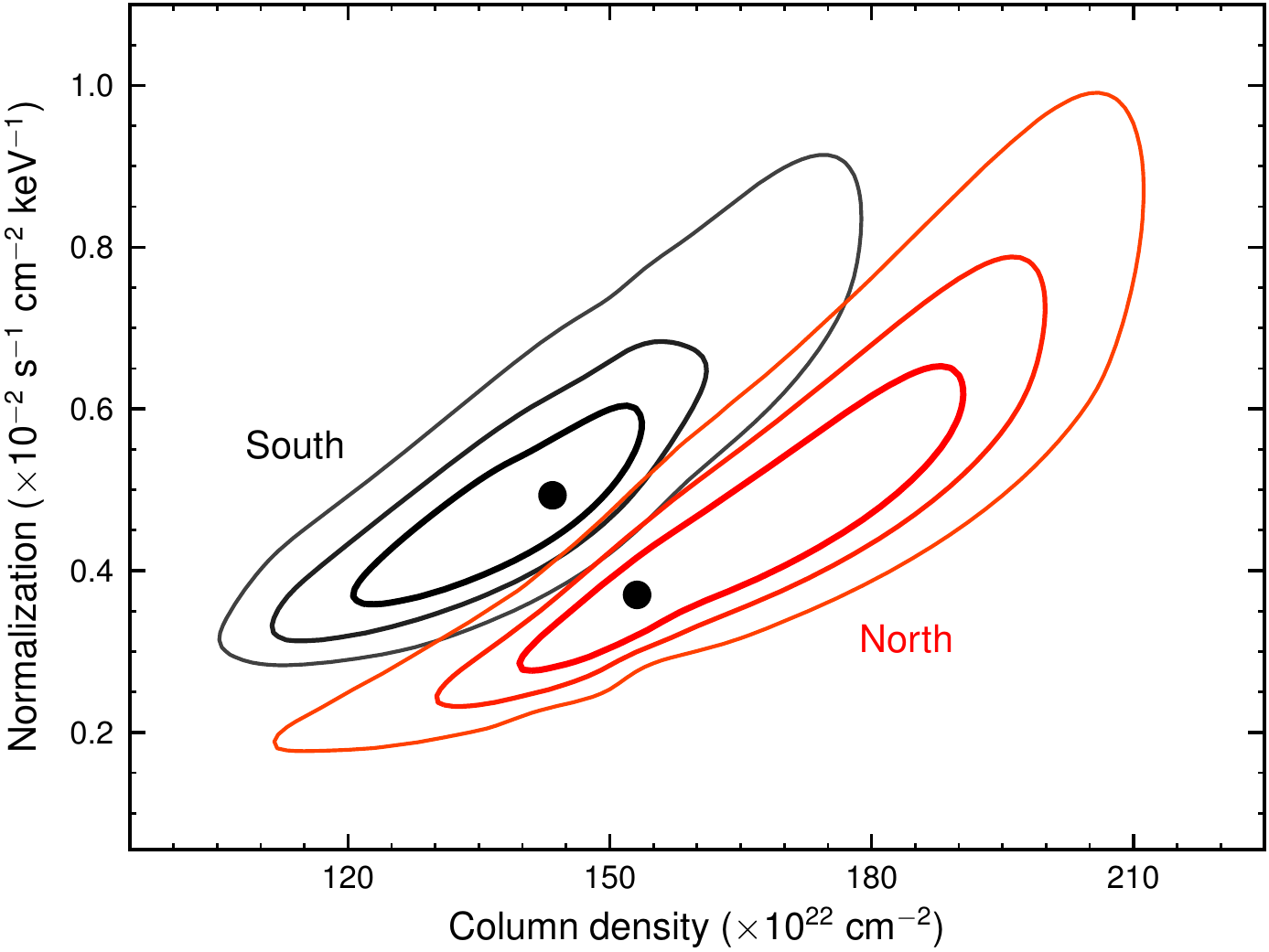}
\caption{Confidence contours at the 68, 90, and 99 per cent levels in the primary 
continuum normalization versus column density plane for both nuclei. Despite the mild 
degeneracy between these quantities, the southern nucleus is generally confirmed to be 
the brighter and less obscured one. For larger differences between the column density 
values, however, solutions of equal intrinsic luminosity are still completely allowed.}
\label{nc}
\end{figure}

For these reasons, simply rescaling the emission measures of the thermal components and 
the normalization of the reflected continuum with respect to the fit of the whole galaxy 
with Model $\mathcal{B}$ results in a poor outcome ($\redchi \sim 1.7$). The column densities 
of the reflected continuum and of the shocked and warm gas were then allowed to vary 
independently for each nucleus, alongside the temperature of the latter. Temperature and
ionization time-scale of the shock were kept frozen to the best-fitting values of Table~\ref{tf}
because unconstrained,\footnote{The fit returns $kT \sim 5.5$ keV and $\tau_\rmn{ion} 
\sim 9 \times 10^{11}$ s cm$^{-3}$, but no useful confidence range can be derived for 
either quantity.} and so were the main parameters of 
the cold gas apart from its emission measure (Model $\mathcal{B}^\prime$). This leads to 
a very good fit, yet the uncertainties on both sets of nuclear variables remain extremely 
large. Following again Fig.~\ref{sf}, it is clear that the AGN-related components become 
rapidly dominant above 10 keV. The average 15--78 keV \nustar spectra were then 
included in the fit to better constrain the primary photon index (tied between the nuclei), 
the intrinsic continuum levels, and the column densities. By construction, this is 
equivalent to imposing that the sum of the resolved 0.5--8 keV spectra smoothly 
connects to the total 15--78 keV emission of the source.\footnote{The same cross 
normalization factors reported in Table~\ref{tf} were assumed here between ACIS-S, FPMA, 
and FPMB.} 
As a result, the re-adjustment of the best-fitting parameters is negligible in each case, 
and an excellent fit is obtained ($\chidof = 428/434$). The statistic further improves by 
$\dchidnu = -13/-$2 if the redshift of the two \sphere components (determined in practice 
by the peak of the \feka feature) is slightly different from the systemic one. This would 
correspond to a blueshift velocity of $-1350^{+660}_{-620}$ \kms for the northern nucleus 
and $-590^{+670}_{-600}$ \kms for the southern one. These shifts were then accepted, even 
though they are well within the energy resolution of the spectra and might not have a real 
physical meaning. 

The final fit results (summarized in Table~\ref{tn}) largely support and complement the 
qualitative picture drawn above. The emission from the shocked gas is about ten times more 
intense nearby the southern nucleus, and it also more absorbed ($\nh \sim 2 \times 10^{23}$ 
cm$^{-2}$). In both cases, the strong features from He- and H-like silicon and sulphur are 
mostly produced in the warm gas phase, whose temperature and obscuration are subject to 
local variations. The reflected continuum from the northern nucleus is attenuated by a 
larger column density, while the opposite trend with similar columns is found for the shock 
component. Some degeneracy cannot be ruled out, but all the parameters, taken individually, 
have reasonable values. Indeed, the reflection efficiency of the two nuclei is $\sim$\,1 
(south) and 2 per cent (north), and the existence of clumps or filaments with $\nh \sim 
10^{22}$--10$^{23}$ cm$^{-2}$ has been firmly established in \ngc. The southern AGN still 
emerges as the more X-ray luminous of the pair, but by 30 per cent only, i.e. much less than 
what would be inferred from the mid-IR flux ratios. These are deeply affected by the fierce 
star formation activity taking place around the southern nucleus (Egami et al. 2006), which 
is also thought to drive the starburst wind (van der Werf et al. 1993; Ohyama, Yoshida \& 
Takata 2003; Feruglio et al. 2013) that shocks the \fexxv-emitting gas. Even if the exact 
X-ray flux and obscuration of the two AGN likely depend upon several 
factors, among which the local properties of the foreground emitting/absorbing gas, the 
present results demonstrate that their having the same intrinsic luminosity is a viable 
solution. Fig.~\ref{nc} shows the continuum intensity versus column density confidence 
contours. These are rather elongated in both cases, implying that the two quantities are 
not entirely independent of one another. In general, however, the greater the column 
density towards the northern nucleus, the smaller the difference in luminosity between 
the two AGN. 

\subsection{Origin of the variability}

While the resort to the average \nustar spectrum has provided new insights into the 
properties of the single nuclei, the nature of the observed hard X-ray variability 
remains unclear. Before analysing the \nustar spectra from the four different epochs, 
we compared the best-fitting Model $\mathcal{B}^\prime$ for the AGN pair to the broad-band 
X-ray emission from the whole galaxy. The nuclear contribution inferred in the previous 
section is consistent with the \nustar data (extracted over a 75-arcsec radius) down to 
$\sim$\,12 keV, where a smooth excess (with marginal structures at 6--7 keV) starts rising. 
Above 3 keV, an excess with the same shape and slightly smaller intensity is found for 
the \chandra spectrum extracted from the central 3-arcsec region, suggesting that this 
is mostly due to the extended shock component, a significant fraction of which is detected 
between the nuclei and around them out to several arcsec (Fig.~\ref{im}; Wang et al. 
2014). While the nuclei are responsible for the observed variability, it is impossible 
to spectrally resolve them within the spatially unresolved spectra, also because of the 
local differences in the foreground thermal components. 

\begin{figure}
\includegraphics[width=8.5cm]{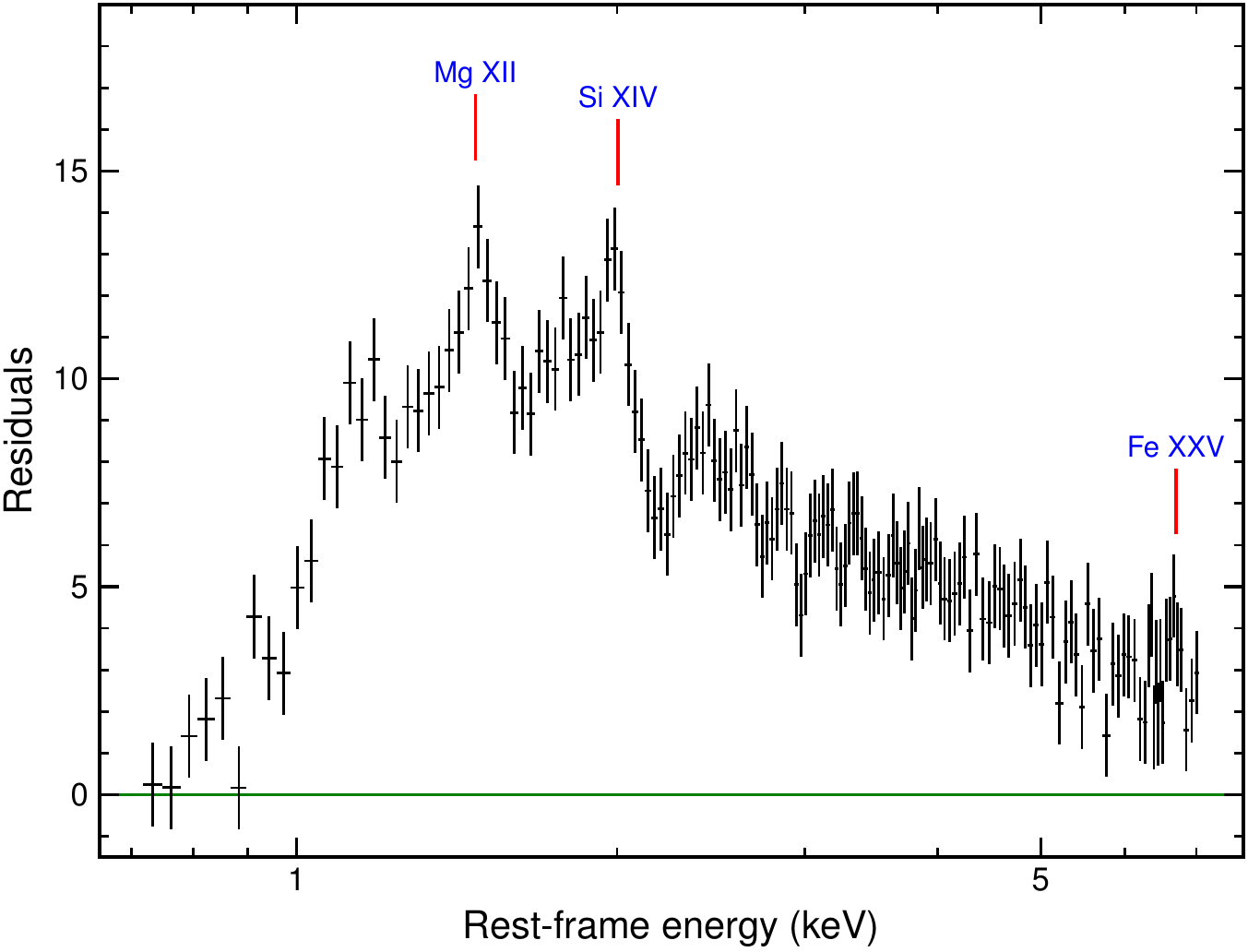}
\caption{Residuals (in units of $\sigma$) in the \chandra spectrum of the whole galaxy 
(rebinned by a factor of two for clarity) with respect to a model consisting of the spatially 
resolved shock, reflected continuum, and nuclear components plus the large-scale emission 
from the cold and warm gas phases. The shape closely resembles a thermal spectrum with an 
absorption cut-off at lower energies. Three major features can be clearly identified from 
Mg\,\textsc{xii}, Si\,\textsc{xiv}, and \fexxv, which mainly originate from the shock-heated 
gas (compare with Fig.~\ref{sf}).}
\label{es}
\end{figure}

To overcome these limitations and explore how the changes in the nuclear properties 
(fluxes and/or column densities) affect the overall variability of \ngc, we thus made 
use of the total \chandra spectrum, freezing all the parameters of the cold and warm gas 
phases to the best-fitting values of Table~\ref{tf}. The emission measure and obscuration 
of the shock component in the neighbourhood of each nucleus, as well as the respective 
reflected continua, were instead fixed following Table~\ref{tn}. This model is expected 
to miss a substantial fraction of the shock emission. Indeed, the residuals (shown in 
Fig.~\ref{es}) are strongly reminiscent of an absorbed, hot thermal component. On these 
grounds, we introduced an additional \texttt{zphabs}$\times$\texttt{pshock} contribution 
with the same temperature and ionization time-scale, while emission measure and obscuration 
were left free to vary. The model so defined was applied to the \chandra and the four 
pairs of \nustar spectra. After verifying that the results are not affected and that the 
cross calibration between FPMA and FPMB is always within 2 per cent (see Table~\ref{to}), 
the merged spectra from both modules were used for ease of data handling, and the \chandra 
data were tied to those from the first \nustar observation. 

\begin{figure}
\includegraphics[width=8.5cm]{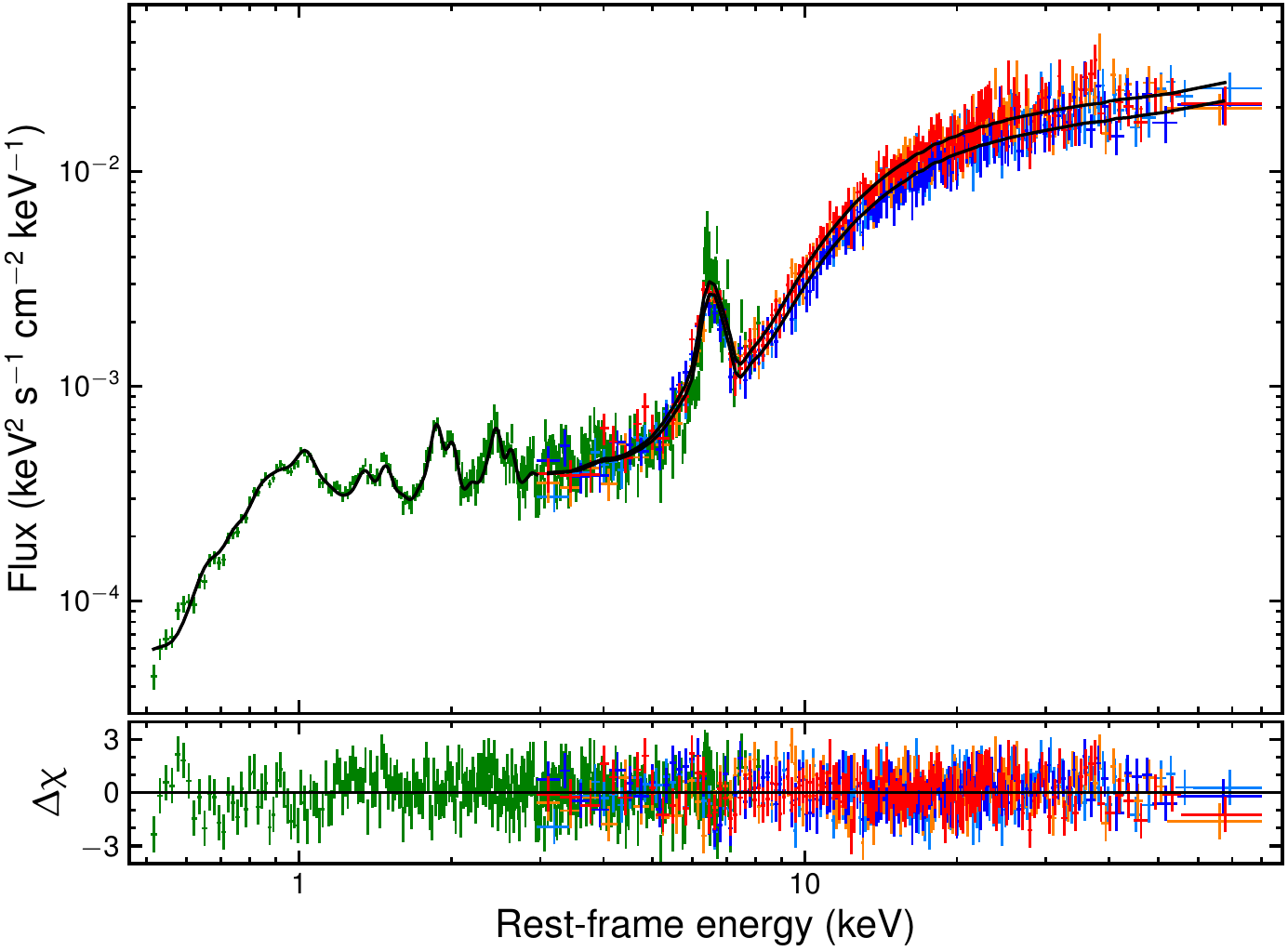}
\caption{Fit to the \chandra and individual \nustar spectra from the four 
observations, with the nuclear components disentangled following the 
spatially-resolved analysis (Table~\ref{tn}) and accounting for the 
extended shock emission (Fig.~\ref{es}). The variability is here ascribed 
to the southern nucleus only. The model divergence above 3 keV between the 
highest (red, 2015 April) and lowest (blue, 2016 February) \nustar flux states 
is shown, with the residuals for all the data sets plotted in the bottom panel.}
\label{mf}
\end{figure}

\begin{figure}
\includegraphics[width=8.5cm]{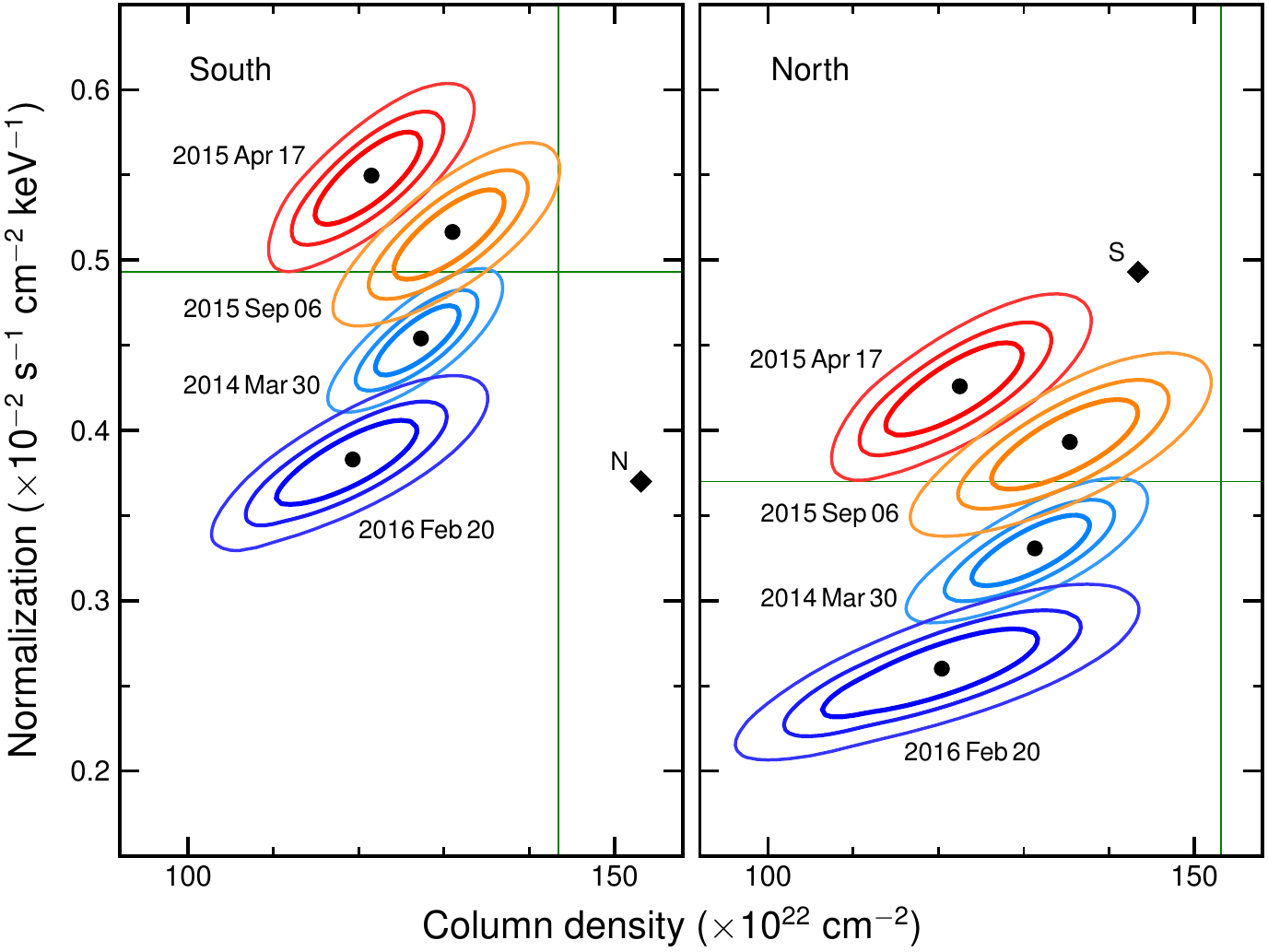}
\caption{Confidence contours as in Fig.~\ref{nc} for a hard X-ray variability entirely 
due to the southern (left) or northern AGN (right). No changes in the column densities 
are statistically required. The cross-hair indicates the position of the best-fitting 
flux and $\nh$ values obtained for that given nucleus from the analysis of the 
time-averaged \nustar spectrum, and assumed in the opposite case of non variability 
(diamond). The colour code for the single observations is the same of Figs.~\ref{lc} 
and \ref{mf}.}
\label{vc}
\end{figure}

Despite the small number of variables (also photon index and iron abundance are 
fixed at this stage), 
the fit is remarkably good, returning a diffuse shock emission with EM $= 95\,(\pm 3) 
\times 10^{63}$ cm$^{-3}$ absorbed by a screen with $\nh 
\simeq 9 \times 10^{21}$ cm$^{-2}$ (compare with Table~\ref{tf}). As the behaviour of the 
two AGN cannot be distinguished, we assumed in turn that one of them remains constant, 
anchoring its flux and column density 
to the time-averaged values of Table~\ref{tn}. All of the variability is therefore entirely 
attributed to the other nucleus. In either case, the fit is fully acceptable, with $\chidof 
\simeq 964/948$ when the northern nucleus varies, and $\chidof \simeq 963/948$ when the 
southern one does. Spectra and residuals of the latter fit are plotted in Fig.~\ref{mf}, 
while the continuum intensity versus column density contours for both nuclei are shown 
in Fig.~\ref{vc}. Statistically, $\nh$ is not required to vary ($\dchidnu \simeq 4/3$ in both 
cases when it is forced to be constant). Quite unpredictably, marginally lower (yet 
fully consistent) $\nh$ values than the ones indicated by the time-averaged analysis seem to 
be involved. This might be due to a systematic effect hidden in our simplified 
assumptions. The choice of common parameters (such as the photon index) between the nuclei, 
for instance, is instrumental in mitigating some degeneracies, but also introduces some 
complex cross dependencies. Moreover, given the mild spectral variability (Fig.~\ref{lc}, 
bottom panel), 
the average \nustar spectrum might not strictly represent any real state of the source, 
so the derived quantities should not be taken at face value.

Even considering the uncertainties above, the nuclear columns are very similar 
(they could differ by as low as $\sim$\,10$^{23}$ cm$^{-2}$; Table~\ref{tn}), hence the 
contour patterns of the two AGN in the X-ray luminosity versus obscuration plane are 
virtually the same. Then the flux range in Fig.~\ref{vc} should only set a lower limit 
to the variability amplitude of each nucleus, which is roughly 40 (south) and 60 per cent 
(north). Indeed, 
for similar luminosities a single AGN can dominate the variability only if its 
central black hole is appreciably less massive than the other, thus accreting at a 
proportionally larger rate (e.g. Ponti et al. 2012, and references therein). The 
masses of the two black holes in \ngc are highly uncertain, but they are estimated 
to have the same order of magnitude, 10$^8$ to 10$^9 M_{\sun}$, and to lie within a 
factor of two from each other (Tecza et al. 2000; Engel et al. 2010a; Medling et al. 
2015). Both AGN thus likely contribute to the variability. In 
this view, the tips and dips of the \swift/BAT light curve could correspond to periods of 
simultaneously high and low states of the two AGN, as the relative flux changes would be 
generally non coherent. Continuum intensity variations as those suggested by Fig.~\ref{vc} 
would give rise to visible effects even at 5--10 keV. Hints of this can be tentatively 
seen in the second \nustar observation (Figs.~\ref{lc} and \ref{mf}). For $\nh < 2 \times 
10^{24}$ cm$^{-2}$, in fact, the direct component is expected to compare with or exceed 
the scattered one, whose fluctuations are reduced by the superposition of different 
reprocessing time 
lags. \chandra and \xmm might have sampled up to seven states 
of \ngc (neglecting the observations separated by just a few days), but for various 
reasons most of them are of little use. Multi-epoch broad-band observations affording 
high quality data below 10 keV, possibly at high spatial resolution, and catching more 
extreme spectral states above 10 keV than the ones probed with this campaign, would be 
therefore needed to achieve a complete X-ray characterization of the two AGN.

\section{Summary and Conclusions}

In this work, we have reported on the analysis of four \nustar observations of the 
galaxy merger \ngc, known for hosting a pair of Compton-thick AGN at a separation 
of less than 1 kpc and a powerful circumnuclear starburst. In the soft X-rays, the 
spectrum of \ngc is due to gas at different temperatures and physical scales. The 
highly obscured nuclei emerge above 4 keV, but the emission from shock-heated gas 
in the starburst superwind is still significant in the \fek band. Although no clear 
variability has ever been 
noticed up to 10 keV, the \swift/BAT light curve shows changes by a factor of two, 
mostly associated with the 14--24 keV energy range (Soldi et al. 2014). The present 
\nustar campaign does not probe any extreme flux states, thus providing a combined 
average spectrum of very good quality. This is best fitted allowing for a nearly 
spherical covering of the X-ray source and slightly subsolar iron abundance 
($Z_\rmn{Fe} \simeq 0.7$--0.8) in the absorber. 

These results have some important implications for the X-ray study 
of the local Compton-thick AGN population. Indeed, the most sophisticated and widely 
used codes of reprocessing by Compton-thick material (e.g. Murphy \& Yaqoob 2009; 
Brightman \& Nandra 2011), based on Monte Carlo simulations, were developed before 
the advent of \nustar. Now that hard X-ray spectra of unprecedented quality have 
become available for tens of highly obscured AGN, the theoretical effort is perhaps 
lagging behind the observations. Especially for the brightest objects, the models 
might lack some flexibility to explore a broader region of the parameter space. It 
would be useful, for instance, to understand the impact of the global covering factor 
and of iron abundance on the measure of the column density, and hence on the estimate 
of the intrinsic AGN luminosity. This is a crucial step to determine the average 
properties of the fainter sources and to assess the contribution of Compton-thick 
AGN to the cosmic X-ray background (e.g. Gilli, Comastri \& Hasinger 2007; Harrison 
et al. 2016), as metallicity was lower at higher redshift. 

The 20 per cent variability observed between the four \nustar observations can be 
entirely explained by changes in the direct continuum flux of the two AGN. When 
their spectra are disentangled below 10 keV thanks to the spatial resolution of 
\chandra, the column densities are quite similar, $\nh \sim 1.2$--$1.5 \times 
10^{24}$ cm$^{-2}$, with the northern nucleus that is generally implied to be 
slightly fainter and more obscured. The larger the difference between the actual 
foreground columns, the smaller the difference in terms of intrinsic AGN luminosity. 
The average absorption-corrected 2--10 keV luminosities are about 1.7 and $2.3 
\times 10^{43}$ \lumcgs for the northern and the southern nucleus, respectively. 
We note, however, that the best-fitting values for both $\nh$ and the primary X-ray 
continuum intensity could be somewhat underestimated due to possible systematics in 
the adopted reprocessing model. The column densities are then likely to fall in the 
range 1.5--$1.9 \times 10^{24}$ cm$^{-2}$, for a cumulative 2--10 keV luminosity 
from the two nuclei of $\sim$\,$6 \times 10^{43}$ \lumcgs. Assuming the standard 
bolometric corrections for obscured AGN (e.g. Lusso et al. 2012; Brightman et al. 
2017), this would suggest that the AGN pair contributes to at least 30 per cent 
of the total energy output of the system. 

Our analysis also provides interesting clues on the nature of X-ray absorption in \ngc. 
The comparable column densities towards the two nuclei, the tentative indication of 
large covering factors, and the lack of evidence for $\nh$ changes jointly point to 
a nuclear obscuration originated by the large amount of gas that is funnelled into 
the central regions following the tidal perturbations at work during the merger 
(Springel, Di Matteo \& Hernquist 2005). The same gas also serves as a reservoir to 
efficiently fuel the starburst activity and the enhanced SMBH accretion. Variations 
of $\nh$ are ubiquitous in isolated AGN, but they are typical of a circumnuclear medium 
that might be clumpy at all scales, like in the revised AGN unification scheme (e.g. 
Netzer 2015), but preserves an ordered structure overall. The nuclear environment 
of \ngc is instead highly disturbed and chaotic, and it is reasonable that both AGN 
are almost completely enshrouded by a nearly uniform shell of dust and gas. Indeed, 
the incidence of fully covered and Compton-thick AGN is known to increase dramatically 
among late-stage mergers (Nardini \& Risaliti 2011; Ricci et al. 2017). If the bulk 
of the obscuring matter is supplied by the merger-driven instabilities, column densities 
of the same order of magnitude for the two nuclei would not be surprising, given the 
similar gas content and bulge mass of the progenitors. 

On the other hand, X-ray 
absorption might be connected to the excited and high-velocity gas in the nuclear 
starburst wind, and inflows/outflows in proximity of the two SMBHs could be the 
cause of the apparent Compton-thickness of the AGN pair. In any case, being one 
of the most spectacular and complex nearby mergers, caught in a relatively rare 
evolutionary stage, \ngc stands out as a unique laboratory to study in detail the 
physics of a prevalent phenomenon of the early Universe, which is the trigger to 
the most luminous episodes of star formation and nuclear activity (e.g. Engel et 
al. 2010b; Treister et al. 2012). 

\section*{Acknowledgments}

The anonymous referee provided useful comments that helped in 
clarifying some parts of the paper. I would also like to thank Tahir Yaqoob 
for useful discussion on \mytorus and on the 
self-consistency issues with the most popular models of transmission and scattering 
in the Compton-thick regime. I acknowledge funding from the European Union's Horizon 
2020 research and innovation programme under the Marie Sk\l{}odowska-Curie grant 
agreement No. 664931. This work was finalized during a visit at the Centre for 
Extragalactic Astronomy, Durham University, to which I am grateful for hospitality. 
The results presented in this paper are based on data obtained with the \nustar 
mission, a project led by the California Institute of Technology, managed by the 
Jet Propulsion Laboratory, and funded by NASA; \xmm, an ESA science mission with 
instruments and contributions directly funded by ESA member states and NASA; and 
the \chandra \textit{X-ray Observatory}. This research has made use of the \nustar 
Data Analysis Software (NuSTARDAS), jointly developed by the ASI Science Data 
Center (Italy) and the California Institute of Technology (USA), and of software 
provided by the \chandra \textit{X-ray Center} (CXC) in the application packages 
\ciao and \sherpa. The figures were generated using SAOImage DS9, developed by SAO, 
and \matplotlib (Hunter 2007), a \python library for publication of quality graphics.



\label{lastpage}

\end{document}